# MyoFold: rapid Myocardial tissue and movement quantification via a highly Folded sequence

Rui Guo[1], Yingwei Fan[1], Bowei Liu[2], Xiaofeng Qian[1], Jiahuan Dai[1], Dongyue Si[2], Yuanyuan Wang[3], Ancong Wang[1], Xiaoying Tang[1*] and Haiyan Ding[2*]

[1]School of Medical Technology, Beijing Institute of Technology, Beijing, China

[2]Center for Biomedical Imaging Research, Department of Biomedical Engineering, School of Medicine, Tsinghua University, Beijing, China

[3]School of Optics and Photonics, Beijing Institute of Technology, Beijing, China

**Correspondence**: Xiaoying Tang[1] and Haiyan Ding[2]

E-mail: xiaoying@bit.edu.cn and dinghy@mail.tsinghua.edu.cn

**Funding:**

National Natural Science Foundation of China for Young Scholars (No. 82202138)

Beijing Institute of Technology Research Fund Program for Young Scholars (No. XSQD-202213003)

The Fundamental Research Funds for the Central Universities (No. LY2022-22)

**Competing interests**

There is a pending patent application for MyoFold. The authors declare that they have no other competing interests.

**Abstract**

**Purpose**: To develop and evaluate a cardiovascular magnetic resonance sequence (MyoFold) for rapid myocardial tissue and movement characterization.

**Method:**

MyoFold sequentially performs joint $T_1/T_2$ mapping and cine for one left-ventricle slice within a breathing-holding of 12 heartbeats. MyoFold uses balanced Steady-State-Free-Precession (bSSFP) with 2-fold acceleration for data readout and adopts an electrocardiogram (ECG) to synchronize the cardiac cycle. MyoFold first acquires six single-shot inversion-recovery images at the diastole of the first six heartbeats. For joint $T_1/T_2$ mapping, $T_2$ preparation ($T_2$-prep) adds different $T_2$ weightings to the last three images. On the remaining six heartbeats, segmented bSSFP is continuously performed for each cardiac phase for cine. We build a neural network and trained it using the numerical simulation of MyoFold for $T_1$ and $T_2$ calculations. MyoFold was validated through phantom and in-vivo experiments and compared to MOLLI, SASHA, $T_2$-prep bSSFP, and convention cine.

**Results:** MyoFold phantom $T_1$ had a 10% overestimation while MyoFold $T_2$ had high accuracy. MyoFold in-vivo $T_1$ had comparable accuracy to that of SASHA and precision to that of MOLLI. MyoFold had good agreement with $T_2$-prep bSSFP in myocardium $T_2$ measurement. There was no difference in the myocardium thickness measurement between the MyoFold cine and convention cine.

**Conclusion:**

MyoFold can simultaneously quantify myocardial tissue and movement, with accuracy and precision comparable to dedicated sequences, saving three-fold scan time.

**Introduction**

Cardiovascular magnetic resonance (CMR) has established various techniques for heart morphology, function, and viability examination. Some of these techniques are considered the gold standard in clinical practice [1]. CMR cine is a method mostly used for capturing the heart contraction over the entire cardiac cycle to assess the function, including wall motion, ejection fraction, mass, thickness, etc. CMR $T_1$-weighted and $T_2$-weighted imaging with contrast or dark-blood technique are used to examine the regional injury and morphology. These qualitative techniques are limited to subjective and global pathologies since there is no reference [2]. CMR parametric mapping through measuring tissue MR relaxation times at a specific magnetic field with and without contrast enables quantitively assessing the pathological deposition of fibrosis, edema, fat, etc., and changes in the extracellular space, providing non-invasive tools for sub-clinical pathology and diffuse alterations [1,3]. Most heart diseases change the heart function, morphology, tissue composition and flow at the same time, such as hypertrophic or dilated cardiomyopathies. The accurate diagnosis and prognosis for these diseases require a comprehensive CMR examination to examine the changes in all these qualitative and quantitative metrics [1,3].

In routine clinical CMR imaging, Modified Look-Locker inversion recovery (MOLLI), $T_2$-prepared balanced Steady-State Free Precession ($T_2$-prep bSSFP), and ECG-triggered segmented bSSFP are used for $T_1$, $T_2$, and heart function examinations [4,5]. These sequences image one LV slice per breathing-holding scan. Following each breathing-holding scan, a rest period, which is almost equal to the scan duration, is required for the patient, resulting in ~50% scan efficiency. Therefore, the inherently extremely long CMR is prolonged with the need for quantification of different tissue properties (i.e. $T_1$, $T_2$, $T_2^*$, etc.) and spatial coverage, increasing the risk of patient noncooperation and overall cost. Besides, images from different breathing-holding scans would not be spatially matched well, impacting the interpretation and limiting other applications, such as image synthesis[6,7].

Alternately, techniques in a multi-tasking fashion have been developed, which have the advantage of saving imaging time and potentially improving the diagnosis through the inherently co-registered tissue MR properties [8]. Most approaches are designed for simultaneous myocardial $T_1$ and $T_2$ mapping [9-16]. Within a single breath-holding scan, these sequences collect ten to twelve hybrid $T_1/T_2$-weighted images to make sure $T_1$ and $T_2$ could be robustly fitted. Cine is needed to be additionally performed for LV function assessment. A few sequences add $T_1/T_2$ weighting during to cine images or use MR Fingerprinting (MRF) or a Multi-tasking scheme, to obtain myocardial $T_1$, $T_2$, and movement simultaneously[16-20]. Although these approaches could utilize the spatial-temporal correction among the signals from different tasking to shorten the imaging time and improve image quality. However, complex and time-consuming reconstruction with the incorporation of physical variations (e.g. heart rate, cardiac motion), imaging parameters (e.g. generating dictionary for MRF reconstruction), or utilizing the low-rank property in images (i.e. low-rank tensor reconstruction for Multitasking) is needed for each scan, [21] [16-18,22], hampering the clinical application. Besides, cine images would be varied by the $T_1/T_2$ preparation, such as these inverted samples, potentially impacting the function quantification or missing several cardiac phases [19,20,23]. Additionally, some sequences have a low temporal resolution for cine [19,20], inducing errors in heart motion estimation. Therefore, there is still a need to develop the clinically feasible sequence with the ability to rapidly offer comprehensive heart examinations.

In this study, we developed a novel CMR sequence, referred to as MyoFold, for rapid quantification of myocardial tissue and heart function. MyoFold is designed to perform $T_1$ mapping, T$_2$ mapping, and cine within a single breath-holding scan of twelve heartbeats. MyoFold highly compresses the joint T$_1$/T$_2$ mapping into the first six heartbeats with a deep-learning fitting method for map building and therefore performs cine subsequently to avoid cross-talk effects between them. MyoFold was validated by phantom and healthy volunteer experiments and compared to MOLLI, T$_2$-prep SSFP, and the conventional ECG-triggered segmented cine.

## Methods

*MyoFold sequence*

**Figure 1A** shows the sequence design of MyoFold, MyoFold sequentially performs joint T$_1$/T$_2$ mapping and cine within a single breath-holding scan of 12 heartbeats. ECG was used to trigger the acquisition of joint T$_1$/T$_2$-weighted images. At the beginning of MyoFold, an inversion pulse is performed, and after a short delay, six single-shot images are collected along the recovery curve of the inverted longitudinal magnetization. Each image is acquired on the diastole of one heartbeat. We added a T$_2$ preparation (T$_2$-prep) pulse before the 4$^{th}$, 5$^{th,}$ and 6$^{th}$ images and changed the echo time of T$_2$-prep to induce different T$_2$ weightings (**Figure 1B**). After the 7$^{th}$ R-wave, segmented bSSFP is performed for each cardiac phase, with a temporal resolution of ~40 ms and ECG is used to synchronize the cardiac cycle. MyoFold cine acquisition is completed on 5-6 heartbeats. The Mz evolution over the data acquisition is shown in **Figure 1C**. The relative contrast between two different tissue is constant during the cine imaging. A 1D neural network, referred to as DeepFittingNet, is used to predict paired T$_1$ and T$_2$ at each pixel [24].

*MR studies*

MyoFold was implemented on a Philips 3T scanner (Achieva TX, Philips Healthcare, Best, Netherlands). A 32-channel cardiac coil was used in in-vivo studies and an eight-channel head MR coil was used in the phantom experiments.

To test the feasibility of MyoFold for in-vivo scans, a total of 21 healthy volunteers (11 males, Age: 26±2 yrs) were recruited. Each subject was scanned by MOLLI, SASHA, T$_2$-prep bSSFP, the conventional segmented cine, and MyoFold under breathing holding with approval from the Institutional Review Board of Tsinghua University. All participants provided written informed consent before MR imaging. For each subject, segmented Cine scans 10-12 LV slices from apex to base in short axis view (SAX), to cover the entire ventricle. Each mapping sequence scans three LV SAX slices at basal, mid, and apical cavities. MyoFold and ECG-segmented Cine additionally image two long-axis slices in two- and four-chamber views, respectively.

For MOLLI, a 5s(3s)3s acquisition scheme was used. Recovery times following two adiabatic inversion pulses were 130 ms and 180 ms, respectively. SASHA collected ten images. Except first one, a WET type saturation pulse was performed before each image [25]. Saturation-recovery time ($T_{SAT}$), which is defined as the duration between the pulse and the central K-space, was increased from 130 ms to $T_{SATmax}$, where $T_{SATmax}$ is the maximum available $T_{SAT}$ regarding the subject's heart rate. T$_2$-prep bSSFP acquired three T$_2$-weighted images with echo time ($TE_{PREP}$) of 0 ms, 25 ms, and 45 ms for adiabatic T$_2$-prep pulse [26]. Parameters used in the readout of three sequences were: ECG-triggered single-shot bSSFP readout, 10 ramp-up pulses, FOV = 300×320×mm$^2$, voxel size = 1.7×2.1×8 mm$^3$, reconstructed voxel size = 0.83×0.83×8 mm$^3$, TR/TE/flip-angle = 2.2ms/1.01ms/35°, SENSE acceleration rate = 2, phase encoding number = 76, acquisition window=165 ms, Bandwidth=1076 Hz, partial echo factor=0.85.

Single-tasking Cine acquires 15 lines in each shot for each cardiac phase using bSSFP readout and ECG trigger. Two-fold compressed-SENSE was used to accelerate acquisition. Other used imaging parameters were: FOV = 300×300 mm$^2$, voxel size = 1.5×1.5×8 mm$^3$, reconstruction voxel size = 0.67×0.67×8 mm$^3$, slice gap=0.8 mm, TR/TE/flip-angle = 3.3ms/1.67ms/40°, 15 lines per shot, shots number = 4, temporal resolution = 50 ms, Bandwidth=1838Hz, Half scan factor = 0.625.

Imaging parameters for MyoFold were: FOV = 300×300 mm$^2$, voxel size = 1.8×1.8×8 mm$^3$, reconstruction voxel size = 1.04×1.04×8 mm$^3$, TR/TE/flip-angle = 2.5ms/1.24ms/40°, SENSE acceleration rate = 2, phase encoding number = 84, acquisition window for joint $T_1/T_2$-weighted image $\sim$= 212 ms, bandwidth=1078Hz, partial echo factor=0.625. For cine acquisition, the lines for each cardiac phase per cardiac cycle was 14 and the temporal resolution was 30-35 ms. Shots for cine were 6. The delay time between the inversion pulse and the start of the acquisition of the first image was 30 ms. Echo times for three $T_2$-prep pulses were 25 ms, 35 ms, and 50 ms.

Phantom experiments were performed to evaluate the accuracy and precision of MyoFold in $T_1$ and $T_2$ measurements. We used 15 phantom tubes. The paired $T_1/T_2$ of these phantoms could almost cover the corresponding values of myocardium and blood at 3T. Reference $T_1$ and $T_2$ of these phantom tubes were measured by an inversion-recovery spin-echo (IR-SE) sequence and a Carr-Purcell-Meiboom-Gill multi-echo SE (CPMG-SE) sequence, respectively. IR-SE acquired 14 images with inversion-recovery time from 100 ms to 3000 ms. The parameters for these two sequences were: FOV = 150×137 mm$^2$, voxel size = 2×2×8 mm$^3$, TR/TE = 15s/9ms. In CPMG-SE, the length of echo trains was ten.

MyoFold was first performed at a simulated heart rate of 60 bpm with five repetitions. The imaging parameters of MyoFold sequences were consistent with those for in-vivo studies. We also performed MyoFold at different heart rates (from 40 bpm to 120 bpm), off-resonance frequencies (from -150 Hz to 150 Hz), and flip angles (from 20° to 40°).

*Image reconstruction*

All images were reconstructed by the scanner. Therefore, after the scan, MyoFold cine was available for interpretation. Elastix motion correction algorithm was applied to images of each mapping sequence to align the myocardium [27]. For IR-SE, CPMG-SE, MOLLI, SASHA, and $T_2$-prep bSSFP, the curve-fitting method was used for the $T_1$ and $T_2$ estimation with three-parameter $T_1$ relaxation and two-parameter $T_2$ relaxation models, respectively.

For MyoFold, DeepFittingNet was used to predict $T_1$ and $T_2$ from hybrid $T_1$/$T_2$-weighted signals, Tis, and TEpreps. DeepFittingNet was implemented using Pytorch Library (Facebook, Menlo Park, California, USA), trained and tested on a standard PC equipped with one CPU (intel i9), one NVIDIA RTX3070 graphic processing unit with 8 GB memory, and 32GB RAM.

DeepFittingNet was trained using the Bloch-equation simulation of MyoFold. The mean absolute error of $T_1$ and $T_2$ was used to update DeepFittingNet parameters. Phantom signals of two repetitions and in-vivo images of six subjects (Males: 2; Age: 26±2 yrs) were used in the validation. The phantom $T_1$ and $T_2$ relative error, and mean LV myocardium $T_1$ and $T_2$ were monitored during the training for the best model selection.

*Post-processing and Statistical Analysis*

For each in-vivo $T_1$ or $T_2$ map, the endo- and epicardial contours were carefully delineated for assessing mean and SD for the entire LV myocardium. Additionally, the septum and blood pool were segmented for measuring the mean and SD of $T_1$/$T_2$ at these two interesting regions.

For each subject, three slices of the conventional cine, which matched well with those by MyoFold, were selected. We used CVI42 (v5.9.3, Cardiovascular Imaging, Calgary, Canada) software to automatically detect the systolic and diastolic phases for cine images of each slice. Then the thickness at systolic and diastolic phases was also automatically calculated by CVI42. Bland-Altman analysis was performed to check the agreement in thickness of the LV myocardium. A paired Student's t-test was conducted to compare the difference between the two approaches in measures of LV function and structural parameters.

The circular region of interest (ROI) for each phantom vial was manually delineated. The mean, standard deviation, and coefficient of variation (CV) for $T_1$ and $T_2$ of pixels within each ROI were calculated. The percentage of relative error regarding the reference value was calculated to evaluate the accuracy. We used SD and CV to evaluate the precision.

Elastix motion correction, $T_1$ and $T_2$ fitting, segmentation, and data analysis were performed on MATLAB 2022a (The MathWorks, Natick, MA).

## Results

All in-vivo and phantom experiments were performed successfully. The training for DeepFittingNet was stopped at 2210 epochs since both the relative errors of phantom $T_1$ and $T_2$ (**Supporting Figure S1A**), and the LV myocardium $T_1$ and $T_2$ (**Supporting Figure S1B**) reached a steady state. Phantom $T_1$ and $T_2$ at the $2033^{rd}$ epoch had the lowest relative errors. Therefore, in this study, we used this model.

**Figures 2** show the raw images of MyoFold joint $T_1/T_2$ mapping and cine at two representative phases. Images for another subject are shown in **Supporting Figure S2**. The corresponding $T_1/T_2$ maps of these two subjects are shown in **Figure 3** and **Supporting Figure S3**, respectively.

Visually, all MyoFold images and maps exhibit homogeneity signals across the LV myocardium and clear myocardium-blood boundaries.

In **Figure 4**, MyoFold yielded high-quality $T_1/T_2$ maps, which were comparable to those by MOLLI and $T_2$-prep bSSFP and better than those from SASHA. **Table 1** lists the LV $T_1$ and $T_2$ for each subject within the testing dataset. **Figure 5** shows the agreements in in-vivo $T_1$ and $T_2$ between MyoFold and these single-tasking sequences. MyoFold had good agreement with SASHA in both LV and septal myocardium $T_1$, with a mean difference of 1 ms (P=0.94) and 22 ms (P=0.16), respectively. The LV and septal myocardium $T_1$ by MOLLI were ~350 ms lower than those by SASHA and MyoFold (all P<0.001). Blood $T_1$ by MyoFold was 460 ms and 121 ms higher than that by MOLLI and SASHA (all p<0.001), respectively. The mean difference in LV myocardium, septum, and blood $T_2$ between $T_2$-prep bSSFP and the proposed MyoFold was 3.11 ms (P<0.001), 2.41 ms (P<0.001) and 13.19 ms (P=0.002), respectively.

In supporting **Table S2**, MyoFold in-vivo $T_1$, and $T_2$ had close SD to those of MOLLI $T_1$ (LV: 106±25 ms vs. 111±31 ms, P=0.68; septum: 76±21ms vs. 83±21 ms, P=0.09) and $T_2$-prep bSSFP $T_2$ (LV: 4.3±1 ms vs. 5±1.2 ms; septum: 3.7±0.9 ms vs. 4.3±1.4 ms. All P<0.05). SASHA in-vivo $T_1$ had the highest SD among the three sequences, with 211±42 ms for LV myocardium and 169±32 ms for septum (all P<0.05 in the student's t-test with both MOLLI and MyoFold). Among the three sequences, MyoFold $T_1$ had the lowest CV (all P<0.05). MyoFold had a higher CV in myocardium $T_2$ measurement than $T_2$-prep bSSFP (all P<0.05). CV of blood $T_2$ by MyoFold was close to that by the conventional cine.

**Figure 6** shows cine images in three different LV planes by the conventional cine and MyoFold. Visually, images from two sequences had clear epi-/endocardial boundaries and good contrast between myocardium and blood. **Figure 7A-D** shows the mean LV myocardium thickness measured by the conventional cine and MyoFold at the systolic and diastolic phases. In **Figure 7E-J**, MyoFold achieved an excellent agreement with the conventional cine in the LV myocardium thickness quantification, confirmed by the results of paired student's t-test (all P>0.05 except P = 0.025 for the middle slice at systolic phase). The mean myocardium thickness by two methods was 10.89±2.07 mm and 11.17±1.83 mm (P = 0.70) at systole, and 6.94±1.17 mm and 6.98±1.11 mm at diastole (P = 0.03).

In phantom experiments (**Figures 8 and Supporting Table S2**), the mean $T_1$ and $T_2$ relative error was -15.79±5.90% for MOLLI (P<0.001), 0.60±5.82% for SASHA (P=0.69), 11.25±5.68% for MyoFold $T_1$ (P<0.001), 19.29±10.08% for $T_2$-prep bSSFP (P<0.001) and 0.24±3.34% for MyoFold $T_2$ (P=0.97), respectively. The mean SD of phantom IR-SE and MyoFold $T_1$ was 23.80±15.88 ms and 21.50 ms ±11.71 ms, and the corresponding mean CV was 1.88±0.60 % (P=0.45) and 1.64±0.46 (P=0.68), respectively. There was no difference in mean SD of phantom $T_2$ between CPMG-SE and MyoFold (0.92±0.33 ms vs. and 0.97 ±0.34 ms, P=0.58), as well as CV (2.33±0.92 % vs. 2.36±0.65%, P=0.90). In **Figure 8 B-D** and **F-H**, the accuracy of MyoFold $T_1$ was increased along with the increase of simulated heart rate, off-center frequency, and flip angle. In contrast, the $T_2$ accuracy was decreased along with these confounders, except for the flip angle.

## Discussion

In this study, we developed and evaluated a highly folded sequence for simultaneous myocardial $T_1$, $T_2$, and heart function examination. In-vivo and phantom validations showed the proposed sequence could measure these quantifications within a single breath-holding scan, with performance comparable to the single-tasking sequences in both accuracy and image quality and three-fold time-saving.

We adopted several strategies to ensure image quality for MyoFold. First, MyoFold is designed to perform cine after the joint $T_1/T_2$ mapping in a separate fashion. Compared to directly adding $T_1$ and $T_2$ preparation to cine continuous acquisition, such a design could avoid the cross-talk effect between them [23]. For example, the myocardium-blood contrast is inverse before and after the nulling points of these two tissues in inversion-recovery $T_1$ mapping, resulting in difficulty in the detection of the endocardium boundary; the cine continuous acquisition would reduce the sensitivity of the hybrid $T_1/T_2$-weighted images to these two parameters, especially $T_2$ weighting [19,20]. Second, MyoFold adopts a deep learning-based approach to map building. Our previous study found that a deep-learning model trained with noise-blurred data for fitting can reduce signal variation [28]. Third, MyoFold used conservative motion compensation manners, that is, breathing holding and ECG trigger. These two methods are mostly used in parametric mapping and cine and exhibit high robustness in image quality. Last, MyoFold employs the inversion-recovery scheme for $T_1$ measurement, which has been demonstrated to have high $T_1$ precision [29].

The clinical utility of CMR has been validated through animal models and patients with various heart diseases. However, the long scan time for an examination of basic heart function and tissue information still hampers its wide applications and therefore increases the subject's cost. In this study, we have demonstrated MyoFold is a simple technique for rapid simultaneous characterization of myocardial $T_1$, $T_2$, and cine within a single breath-holding scan. Compared to the daily used dedicated sequences for these quantifications, MyoFold could reduce the total breathing-holding times and shorten the scan time. Hence, in the future, along with other accelerated CMR sequences, MyoFold is a promising technique to shorten the CMR examination to 30 min or less [30].

MyoFold is a clinically feasible technique. The technique of trajectory, acceleration, and readout adopted in MyoFold are widely available on most vendors and widely used in most mapping and cine sequences. Therefore, all MyoFold images could be reconstructed by the scanner, avoiding long and complex offline processing. DeepFittingNet is a simple full-connected neural network and easy to be implemented on the vender. We have implemented such a neural network on the SIEMENS scanner for in-line $T_1$ map reconstruction in a previous study [31]. In the future, by installing DeepFittingNet on the scanner, all MyoFold maps and cine images could be available for analysis/interpretation after the scan, to further improve the clinical feasibility.

We trained DeepFittingNet using ground truth $T_1$ and $T_2$ of the simulated signals as the references. Therefore, MyoFold exhibited good $T_1$ and $T_2$ accuracy. MyoFold $T_1$ had slight overestimation in the phantom study, while in the in-vivo study, MyoFold $T_1$ was close to that measured by SASHA, the latter had been demonstrated to have higher $T_1$ accuracy [29]. MyoFold $T_2$ had high accuracy in phantom experiments. In-vivo $T_2$ at 3T by MyoFold was slightly lower than that by $T_2$-prep bSSFP, but close to reported values [32,33]. $T_2$-prep bSSFP is known to overestimate $T_2$ values when using a two-parameter fitting model [34,35]. In this study, we used one model to predicate MyoFold $T_1$ and $T_2$ jointly. The accuracy of each parameter would be impacted by another (Figure 8B and F). Further study is warranted to investigate whether two independent models could improve $T_2$ accuracy.

In the future, by using gradient echo, radial trajectory, and mixed single-/multi-echo readout in MyoFold, myocardial $T_2^*$, which is another important quantification for iron overload, could be measured. The joint $T_1/T_2$ images and the heartbeats for the cine could be further optimized to shorten the total imaging time. The imaging parameters, such as flip angle, repeated time, and partial-echo factor, could be optimized to further improve myocardium-blood contrast.

There are several limitations in this study. All participants were young and healthy since our biomedical imaging research center is affiliated to and located in a technological university. Further validations with patients are warranted to evaluate the performance of the proposed MyoFold. MyoFold scan was added to the end of an already ~50min CMR scan. Therefore, we only imaged three slices for each subject by MyoFold. Several quantifications of heart function were not able to be measured. We did not optimize the parameters of DeepFittingNet for the MyoFold $T_1$ and $T_2$ prediction.

**Conclusion**

MyoFold as proposed and validated in this study enables the examination of myocardium $T_1$, $T_2$, and heart function by single a breath-holding scan with performance in accuracy, precision, and image quality comparable to the most used dedicated sequences. MyoFold is a promising technique for shortening the CMR examination. Further optimizations and feasibility studies are warranted to evaluate its clinical utility.

# Figures and Captions

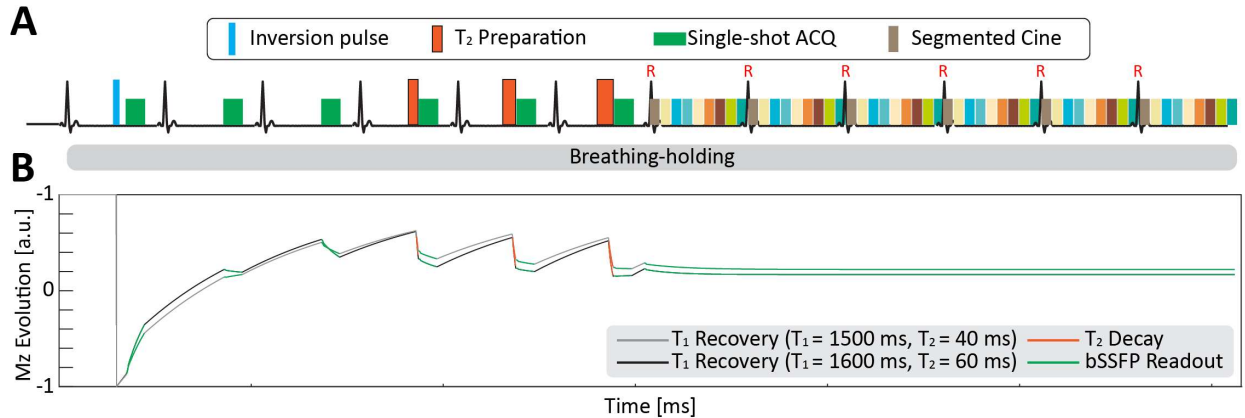

**Figure 1. A:** Breathing-holding MyoFold sequence design. **B:** Evolution of the longitudinal magnetization (Mz) during the readout. MyoFold sequentially performs joint $T_1$/$T_2$ mapping and cine in 12 heartbeats. Balanced Steady-State-Free-Precession (bSSFP) is used for data readout and the electrocardiogram (ECG) is used for cardiac cycle synchronization. MyoFold first acquires six single-shot images. The inversion pulse is performed before the 1st image, and $T_2$ preparation is additionally performed before the readout of the 4th, 5th, and 6th images to induce $T_2$ weighting. Segmented bSSFP is continuously performed on the rest 6 heartbeats for each cardiac phase for cine.

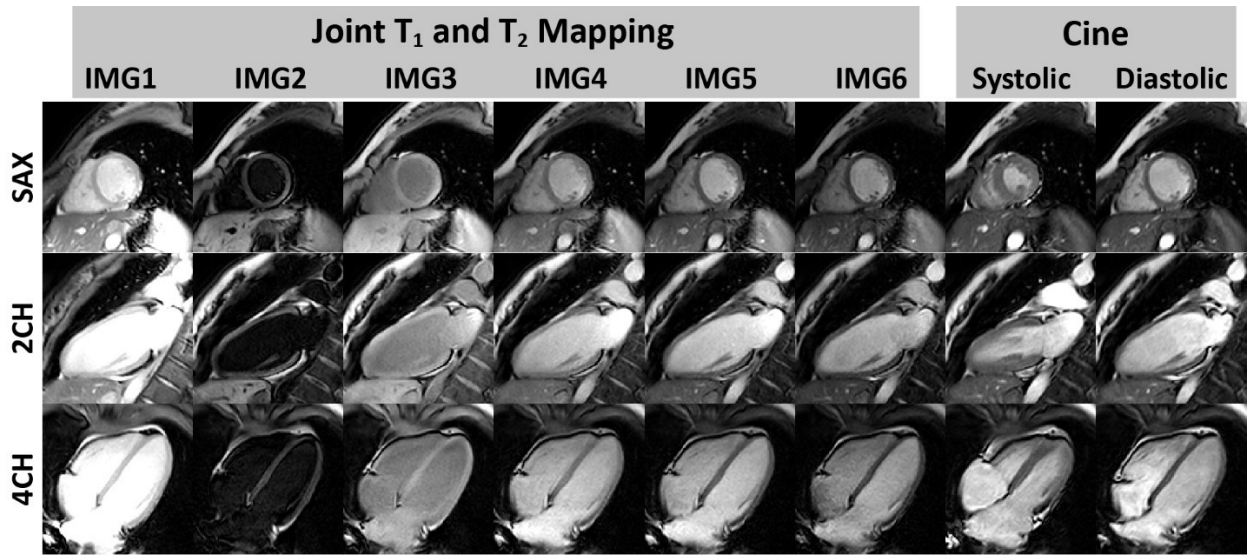

**Figure 2.** MyoFold hybrid $T_1/T_2$-weighted images (IMG#) and cine (Systolic vs. Diastolic) of a volunteer in three views. Corresponding videos are shown in **Supporting Files 2**.

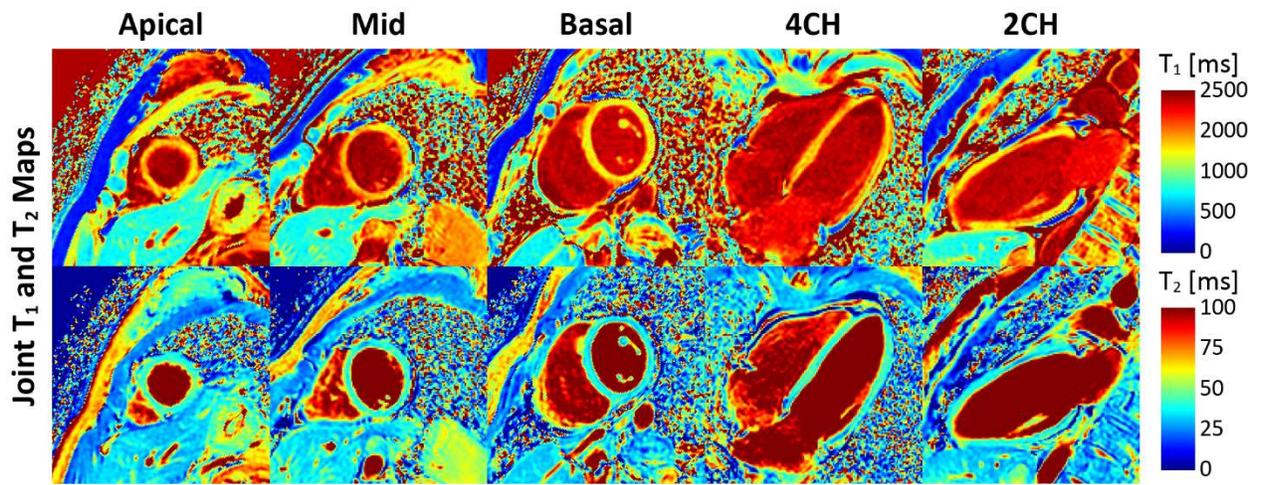

**Figure 3.** MyoFold joint $T_1$ and $T_2$ maps for weighted images in **Figure 2**.

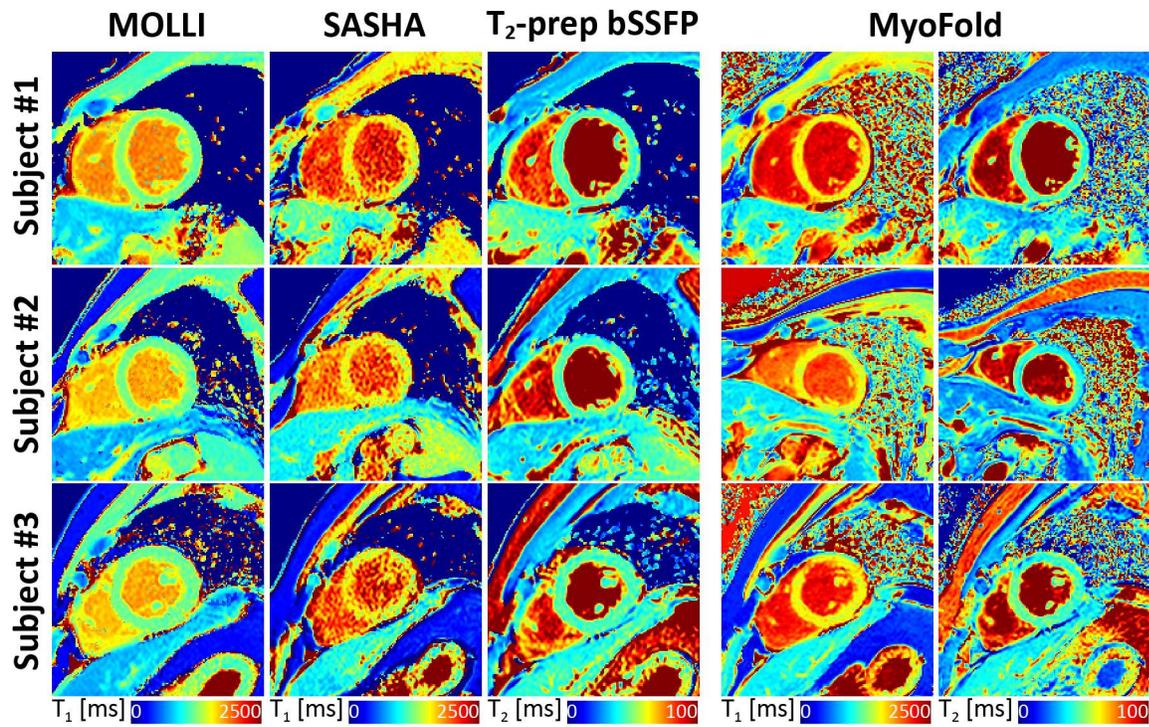

**Figure 4.** Comparing MyoFold joint $T_1$ and $T_2$ maps of four subjects to those by MOLLI, SASHA, and $T_2$-prep bSSFP. Visually, MyoFold MOLLI and $T_2$-prep bSSFP had better image quality than SASHA.

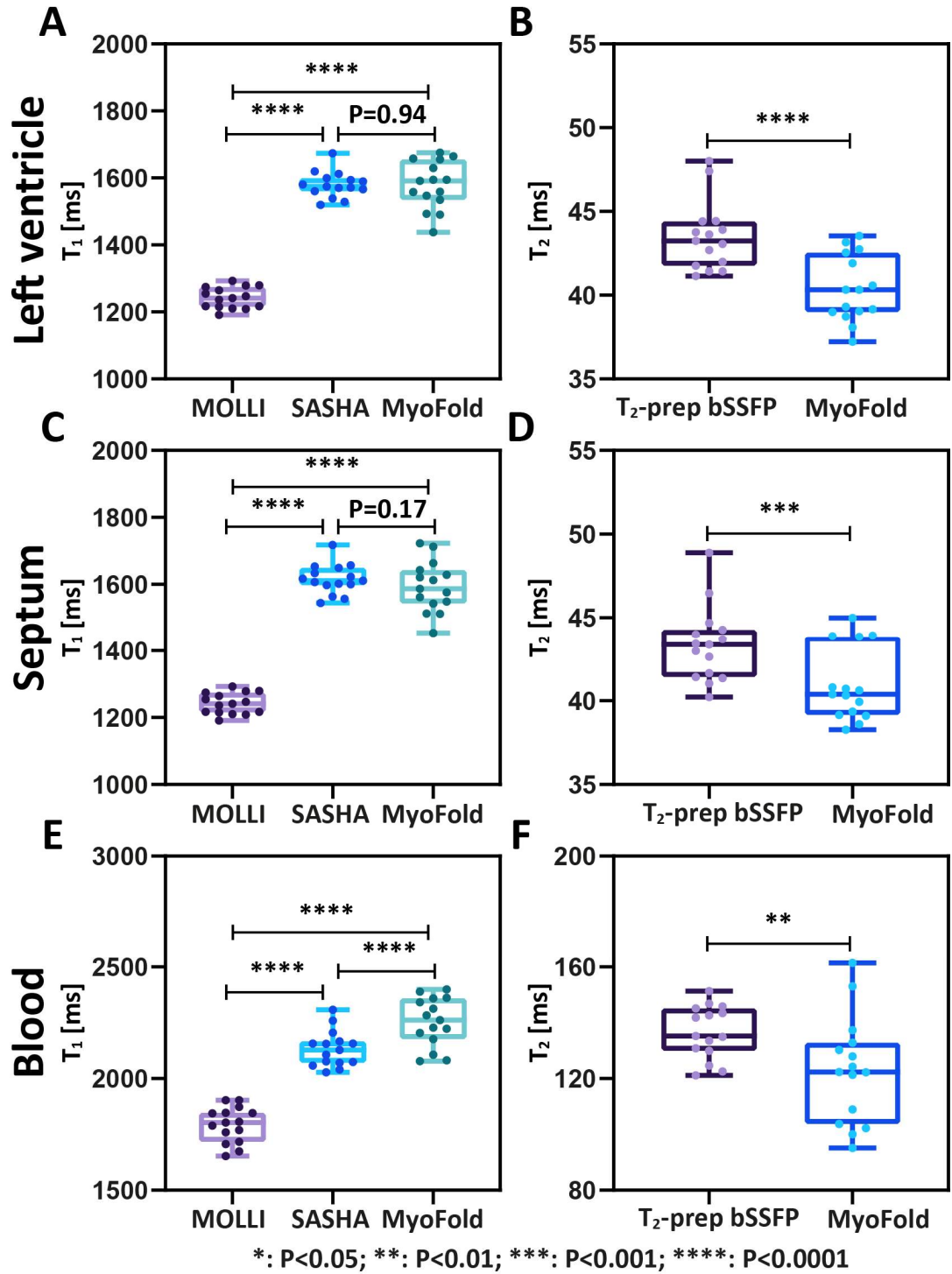

*: P<0.05; **: P<0.01; ***: P<0.001; ****: P<0.0001

**Figure 5.** Left-ventricle myocardium, septum, and blood $T_1$ of each subject measured by three sequences, as well as $T_2$.

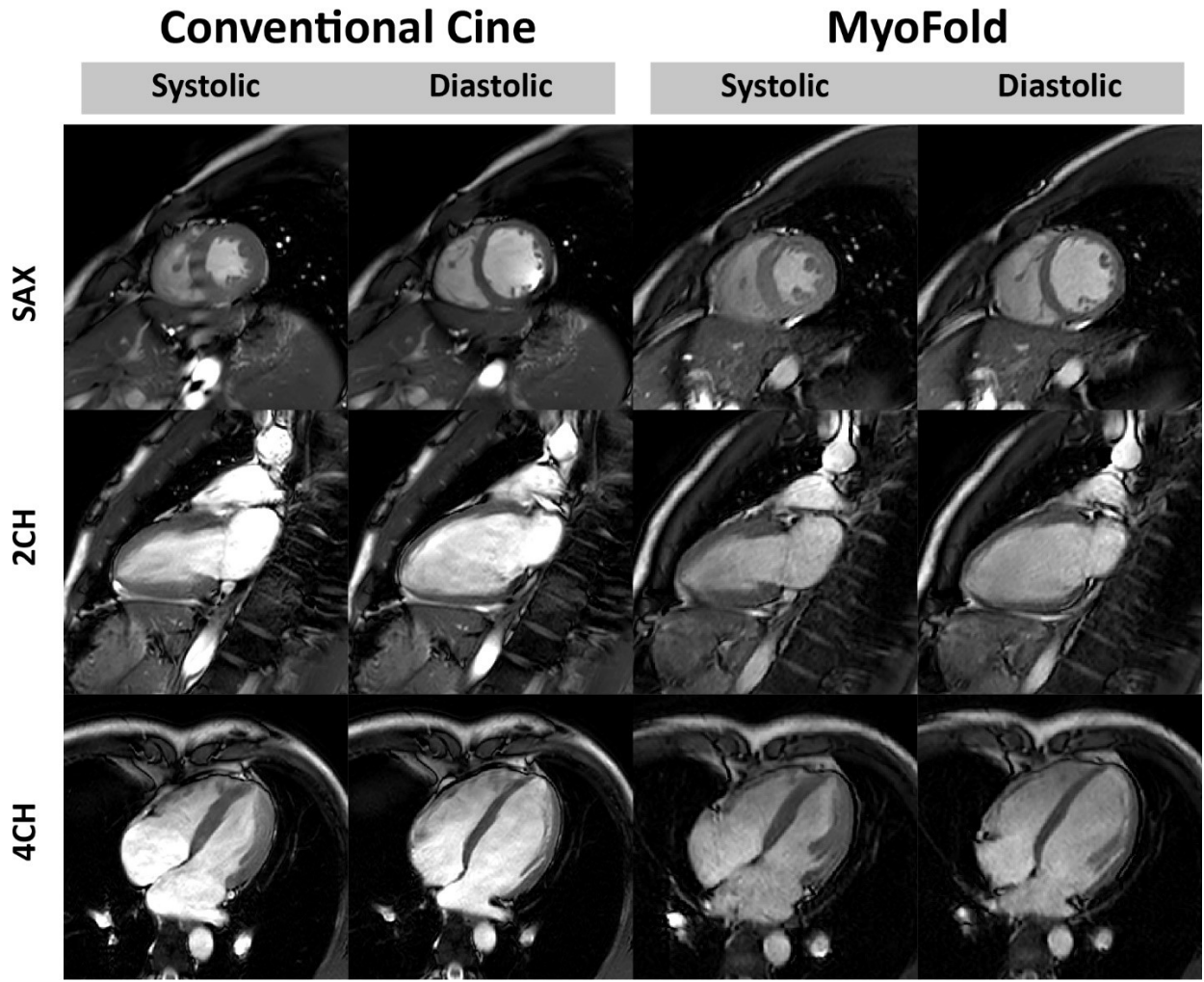

**Figure 6.** Comparing MyoFold to the conventional cine. **Supporting Files 3** and **4** shows the corresponding videos.

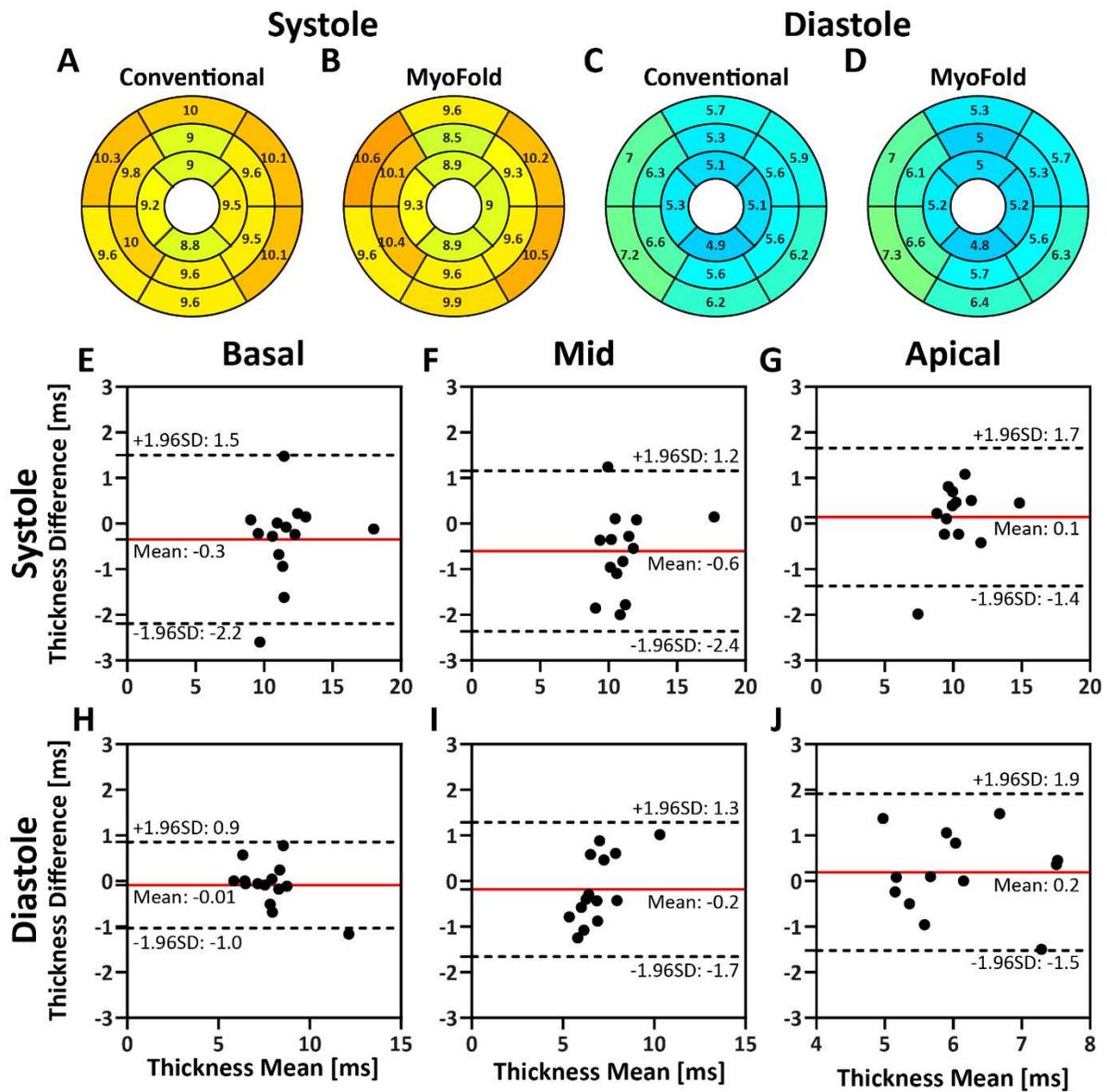

**Figure 7.** Agreements in left-ventricle myocardium thickness between the conventional cine and MyoFold.

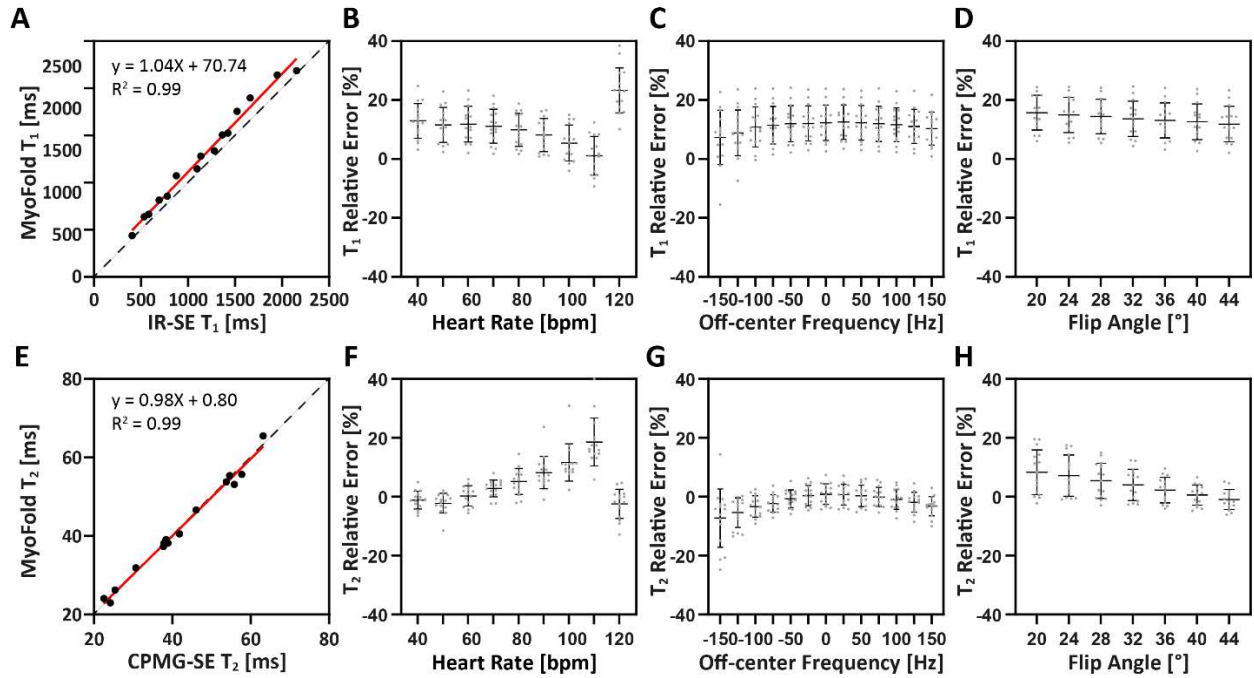

**Figure 8.** Comparing MyoFold phantom $T_1$ and $T_2$ to reference values by spin-echo sequences.

**Table 1.** Characteristics, myocardium $T_1$ and $T_2$ for each subject within the testing dataset.

| ID | Sex | Age | Height | Weight | Heart rate | Left-ventricle $T_1$ | | | | | | Left-ventricle $T_2$ | | | |
|---|---|---|---|---|---|---|---|---|---|---|---|---|---|---|---|
| | | | | | | MOLLI | | SASHA | | MyoFold | | $T_2$-prep bSSFP | | MyoFold | |
| | | | | | | Mean | SD | Mean | SD | Mean | SD | Mean | SD | Mean | SD |
| 1 | M | 25 | 168 | 62 | 60 | 1217 | 97 | 1528 | 169 | 1547 | 79 | 43.9 | 3.7 | 42.5 | 4.6 |
| 2 | M | 29 | 171 | 60 | 69 | 1209 | 79 | 1561 | 178 | 1490 | 71 | 43.8 | 3.7 | 39 | 3.5 |
| 3 | M | 23 | 174 | 82 | 66 | 1215 | 92 | 1594 | 219 | 1459 | 77 | 41.8 | 4 | 39.3 | 3.1 |
| 4 | F | 24 | 160 | 49 | 67 | 1280 | 112 | 1620 | 170 | 1630 | 96 | 44.4 | 3.9 | 40.3 | 4.4 |
| 5 | M | 27 | 178 | 80 | 69 | 1191 | 125 | 1520 | 181 | 1493 | 95 | 41.4 | 3.7 | 38.7 | 4.3 |
| 6 | F | 25 | 166 | 54 | 59 | 1293 | 106 | 1674 | 192 | 1655 | 116 | 47.4 | 4.9 | 42.7 | 5.7 |
| 7 | F | 23 | 162 | 62 | 80 | 1265 | 113 | 1599 | 202 | 1591 | 120 | 42 | 3.7 | 39 | 4.5 |
| 8 | M | 29 | 168 | 84 | 72 | 1236 | 116 | 1612 | 229 | 1658 | 133 | 44.4 | 5.4 | 41.9 | 6.4 |
| 9 | M | 27 | 178 | 80 | 69 | 1247 | 98 | 1539 | 264 | 1559 | 95 | 42.7 | 3.7 | 38.1 | 4.1 |
| 10 | M | 27 | 178 | 86 | 79 | 1217 | 104 | 1575 | 233 | 1558 | 100 | 43.6 | 3.9 | 43.2 | 4.8 |
| 11 | M | 25 | 178 | 67 | 65 | 1241 | 115 | 1590 | 230 | 1594 | 110 | 48 | 5.5 | 43.5 | 5.9 |
| 12 | F | 28 | 170 | 71 | 84 | 1274 | 125 | 1581 | 187 | 1675 | 109 | 43.2 | 4.7 | 40.3 | 4.9 |
| 13 | F | 25 | 155 | 55 | 87 | 1279 | 133 | 1571 | 247 | 1656 | 153 | 41.1 | 4.3 | 41.6 | 6.2 |
| 14 | F | 27 | 158 | 48 | 73 | 1255 | 98 | 1571 | 198 | 1595 | 116 | 43.1 | 3.5 | 39.2 | 5.3 |
| 15 | M | 31 | 160 | 56 | 75 | 1211 | 78 | 1566 | 271 | 1534 | 138 | 41.4 | 5 | 39.3 | 6 |

# Additional File 1

**Supporting Table S1.** The mean and standard deviation (SD) of $T_1$ and $T_2$ for each phantom vial.

| # | IR-SE $T_1$ | | CPMG-SE $T_2$ | | MOLLI | | SASHA | | MyoFold $T_1$ | | $T_2$-prep bSSFP | | MyoFold $T_2$ | |
|---|------|------|------|------|------|------|------|------|------|------|------|------|------|------|
|   | Mean | SD | Mean | SD | Mean | SD | Mean | SD | Mean | SD | Mean | SD | Mean | SD |
| 1 | 405 | 5 | 46.1 | 1.6 | 365 | 5.2 | 411 | 13.3 | 435 | 13 | 67.2 | 2 | 46.6 | 1.4 |
| 2 | 534 | 6.7 | 53.8 | 1 | 499 | 12.5 | 560 | 12.8 | 636 | 13.4 | 70 | 1.2 | 53.8 | 1.1 |
| 3 | 583 | 4.7 | 55.8 | 0.9 | 506 | 22 | 581 | 18.2 | 663 | 14.4 | 68.9 | 2.2 | 53.1 | 1.1 |
| 4 | 694 | 9.3 | 38.9 | 0.9 | 633 | 6 | 721 | 10.3 | 813 | 14.9 | 49.7 | 1.2 | 38.2 | 0.8 |
| 5 | 781 | 16.5 | 41.8 | 1.5 | 657 | 7.8 | 735 | 19.8 | 856 | 14.9 | 51.7 | 1.3 | 40.5 | 1 |
| 6 | 876 | 9.4 | 57.7 | 1.2 | 834 | 8.2 | 902 | 17 | 1072 | 15.9 | 66.3 | 1.2 | 55.7 | 1.2 |
| 7 | 1097 | 22.6 | 24.2 | 0.9 | 859 | 9.5 | 1110 | 25.1 | 1144 | 13.8 | 30 | 0.5 | 23 | 0.6 |
| 8 | 1136 | 18.8 | 37.9 | 0.7 | 981 | 9.3 | 1138 | 27.2 | 1280 | 17.7 | 43.9 | 0.8 | 38.2 | 0.8 |
| 9 | 1280 | 21.2 | 30.7 | 0.4 | 1007 | 8.1 | 1223 | 35.2 | 1336 | 12.1 | 36.2 | 0.4 | 31.9 | 0.6 |
| 10 | 1365 | 28.4 | 54.6 | 0.5 | 1159 | 5.5 | 1300 | 26.7 | 1505 | 13.9 | 60 | 0.5 | 55.4 | 0.7 |
| 11 | 1425 | 43 | 38.4 | 0.8 | 1145 | 11 | 1360 | 51.1 | 1524 | 18.3 | 44.5 | 0.8 | 39.1 | 0.9 |
| 12 | 1521 | 34.8 | 25.4 | 1 | 1204 | 25 | 1695 | 73.7 | 1755 | 36.8 | 28.4 | 0.7 | 26.2 | 0.5 |
| 13 | 1661 | 30.6 | 22.5 | 0.7 | 1312 | 12.2 | 1792 | 44.5 | 1896 | 42.4 | 26.2 | 0.3 | 24.1 | 0.6 |
| 14 | 1950 | 50.5 | 37.7 | 0.6 | 1547 | 23.4 | 1901 | 97.5 | 2140 | 48.5 | 40.3 | 1 | 37.3 | 1.6 |
| 15 | 2155 | 55.5 | 63.2 | 1.1 | 1644 | 13.8 | 1878 | 80.7 | 2186 | 32.6 | 66.2 | 1.5 | 65.5 | 1.5 |

\* IR-SE: Inversion-Recovery Spin-Echo; CPMG-SE: Carr-Purcell-Meiboom-Gill Spin-Echo; MOLLI: Modified Look-Locker inversion recovery. SASHA: saturation recovery single-shot acquisition;
\*\*Results of MOLLI3(3)3(3)5, MOLLI5(3)3, SASHA and $T_2$-prep bSSFP were averaged from two repetitions.

**Supporting Table S2.** Mean, standard deviation, and CV for in-vivo $T_1$ and $T_2$ of all normal subjects (N=14) at 3T

|  |  | $T_1$ | | | $T_2$ | |
| --- | --- | --- | --- | --- | --- | --- |
|  |  | **MOLLI** | **SASHA** | **MyoFold** | **$T_2$-prep bSSFP** | **MyoFold** |
| **Left ventricle** | **Mean [ms]** | 1242±35 | 1580±50 | 1584±75 | 43.5±2.4 | 40.3±2.6 |
|  | **SD [ms]** | 106±25 | 211±42 | 111±31 | 4.3±1 | 5±1.2 |
|  | **CV [%]** | 8.5±2 | 13.4±2.6 | 7±1.8 | 9.8±2.1 | 12.2±2.7 |
| **Blood** | **Mean [ms]** | 1793±86 | 2133±95 | 2258±117 | 136.7±10.9 | 120.8±21.6 |
|  | **SD [ms]** | 183±99 | 218±56 | 59±15 | 16.1±4.3 | 12±4 |
|  | **CV [%]** | 10.3±5.8 | 10.2±2.7 | 2.6±0.7 | 11.7±2.9 | 10.1±3.2 |
| **Septum** | **Mean [ms]** | 1255±41 | 1616±56 | 1598±80 | 43.3±2.8 | 40.9±3.1 |
|  | **SD [ms]** | 76±21 | 169±32 | 83±21 | 3.7±0.9 | 4.3±1.4 |
|  | **CV [%]** | 6.1±1.6 | 10.4±2 | 5.2±1.3 | 8.5±1.9 | 10.6±3.3 |

\*SD: standard deviation; CV: coefficient of variation.

\*\*Each result is averaged from all slices of all subjects.

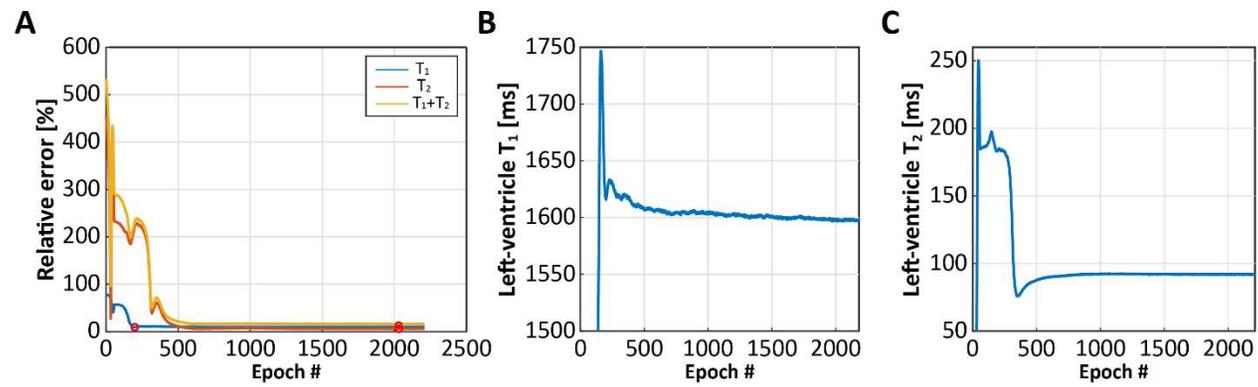

**Support Figure S1.** Results of validation. **A:** The relative error of phantom $T_1$ and $T_2$ over epochs, with spin-echo references. **B** and **C:** left-ventricle $T_1$ and $T_2$ of validation at each epoch.

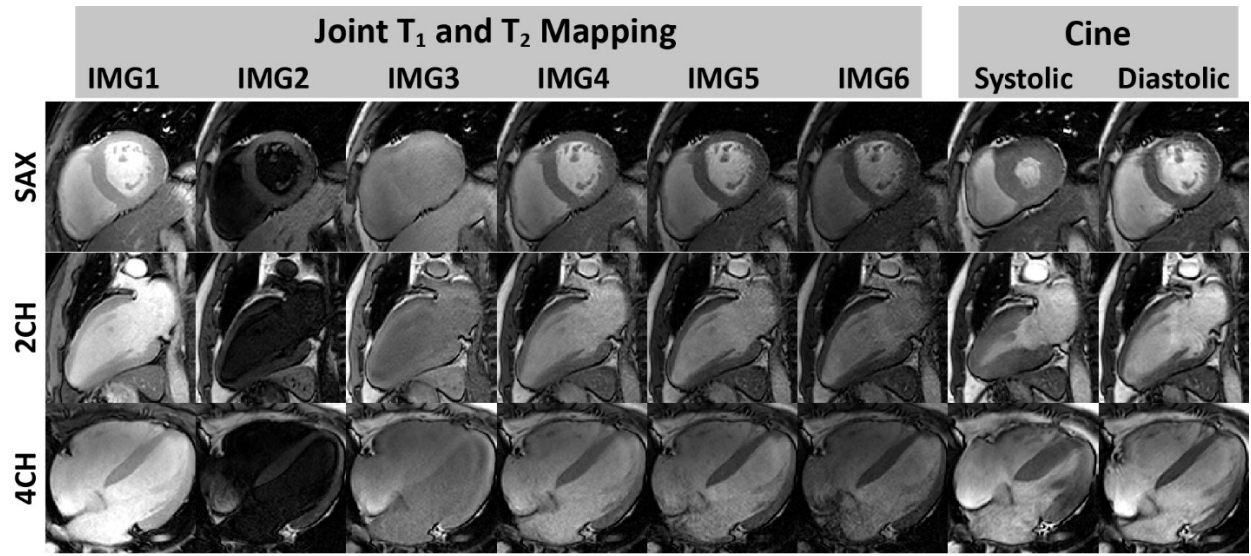

**Support Figure S2.** Hybrid $T_1/T_2$-weighted images (IMG#) and cine (Systolic vs. Diastolic) of one volunteer from three different views. Corresponding videos are shown in **Supporting File S3**.

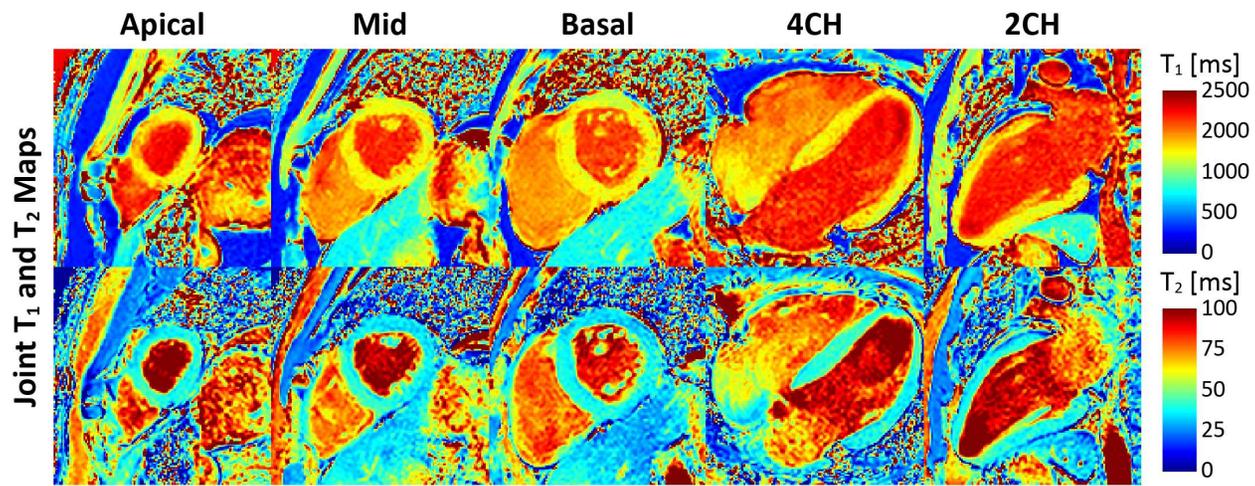

**Support Figure S3.** MyoFold joint $T_1$ and $T_2$ maps for images are shown in **Figure S2**.